\documentclass[aps,prb,preprint]{revtex4-1}
\usepackage{graphicx}
\usepackage{dcolumn}
\usepackage{bm}

\begin{document}

\preprint{}

\title{Strong magnetoelastic effect on the magnetoelectric phenomena of TbMn$_{2}$O$_{5}$}
\author{Yoon Seok Oh,$^{1}$ Byung-Gu Jeon,$^{1}$ S. Y. Haam,$^{1}$ S. Park,$^{2}$ V. F. Correa,$^{3}$ A. H. Lacerda,$^{4}$ S.-W. Cheong,$^{2}$ Gun Sang Jeon,$^{1}$ and Kee Hoon Kim,$^{1,*}$}

\address{$^{1}$CeNSCMR, Department of Physics and Astronomy, Seoul National University, Seoul 151-747, Republic of Korea\\
$^{2}$Rutgers Center for Emergent Materials and Department of Physics $\&$ Astronomy, Piscataway, NJ 08854, USA\\
$^{3}$Comisi\'{o}n Nacional de Energ\'{i}a At\'{o}mica, Centro At\'{o}mico Bariloche, 8400 S. C. de Bariloche, Argentina\\
$^{4}$LANSCE, Los Alamos National Laboratory, Los Alamos, NM 87545, USA
}

\date{\today}

\begin{abstract}
Comparative studies of magnetoelectric susceptibility ($\alpha$), magnetization ($M$), and magnetostriction ($u$) in TbMn$_{2}$O$_{5}$ reveal that the increment of $M$ owing to the field-induced Tb$^{3+}$ spin alignment coins a field-asymmetric line shape in the $\alpha(H)$ curve, being conspicuous in a low temperature incommensurate phase but persistently subsisting in the entire ferroelectric phase. Correlations among electric polarization, $u$, and $M^{2}$ variation represent linear relationships, unambiguously showing the significant role of Tb magnetoelastic effects on the low field magnetoelectric phenomena of TbMn$_{2}$O$_{5}$. An effective free energy capturing the observed experimental features is also suggested.
\end{abstract}


\maketitle

Nontrivial cross-coupling between electric and magnetic dipoles realized in multiferroics has been a subject of extensive researches in recent years, which are targeted to understand the mechanism of magnetoelectric (ME) coupling as well as to find novel device applications. One of the key compounds that has triggered such research activity is TbMn$_{2}$O$_{5}$, in which a continuous actuation of electric polarization ($P$) is realized within low magnetic field ($H$) below 2 T. Numerous studies on this compound and related $R$Mn$_{2}$O$_{5}$ ($R$ = Y, Dy, Ho, Er, and Bi) have shown that spatially modulating, noncollinear magnetic order due to spin frustration is responsible for inducing ferroelectric order in these materials. More specifically, a main mechanism for having nontrivial $P$ in $R$Mn$_{2}$O$_{5}$ ($R$ = Tb, Y, Dy, and Bi) has been attributed to exchange striction among frustrated Mn spin networks,\cite{chapon1,chapon2,kagomiya,blake,jwkim} while $P$ contribution from spiral spin order has also been known to be important in $R$Mn$_{2}$O$_{5}$ ($R$ = Ho, Er, and Tm).\cite{hkimura,mfukunaga} Thus, a main mechanism for developing $P$ in $R$Mn$_{2}$O$_{5}$ can be arguably dependent on a specific material while it is obviously associated with the Mn spin order.\cite{chapon1,chapon2,kagomiya,blake,jwkim,hkimura,mfukunaga,jokamoto,jkoo}

Only a limited number of works have discussed the possible effects of rare earth ions on the temperature- and $H$-dependence of $P$ on $R$Mn$_{2}$O$_{5}$.\cite{jkoo,wratcliff,tyson,lottermoser} As a result, a proper role of rare earth ions on the ME phenomena of $R$Mn$_{2}$O$_{5}$ is far from complete understanding and thus worthy of investigation. One particularly important question is how one can understand the $H$-induced actuation of $P$ that is uniquely realized in TbMn$_{2}$O$_{5}$. A detailed understanding of this intriguing question is likely to provide not only an answer for the long-standing puzzle that has triggered the multiferroic research but also useful information regarding the application of multiferroics.

In this communication, on the basis of systematic studies of magnetostriction ($u$), magnetization ($M$), and ME susceptibility ($\alpha$), we uncover that $M$ change due to Tb spin alignment with $H$ determines the evolution of both $u$ and $P$ predominantly, thereby developing linear relationships among $M^{2}$, $u$, and $P$ in the entire ferroelectric phase. An effective free energy analysis based on the magnetoelastic coupling of Tb can successfully describe the experimentally found correlation among those physical quantities.

Single crystals of TbMn$_{2}$O$_{5}$ were grown with a PbO:PbF$_{2}$ flux.\cite{nhur} To investigate detailed $H$-and temperature-dependent $P$ ($//b$) and lateral length $l$ ($//a$) change, we developed a sensitive ME susceptometer and a high precision dilatometer, both of which operate in a PPMS$^{\rm TM}$. In this study, we have focused on $\alpha_{21}=\delta P_{b}/\delta H_{a}$ and longitudinal magnetostriction $u_{a}\equiv (l(H_{a})-l(0))/l(0)$ along the $a$-axis. For the former, we used solenoid coils to apply small ac $H$ ($//a$) of $\sim$4 Oe and a high impedance charge amplifier to sensitively detect an ac modulated charge, proportional to $\delta P_{b}$, by using a lock-in technique.\cite{hryu} Dielectric constant ($\epsilon$) and $M$ were also investigated with a capacitance bridge and a vibrating sample magnetometer, respectively.

\begin{figure}
\begin{center}
\includegraphics[width=0.48\textwidth]{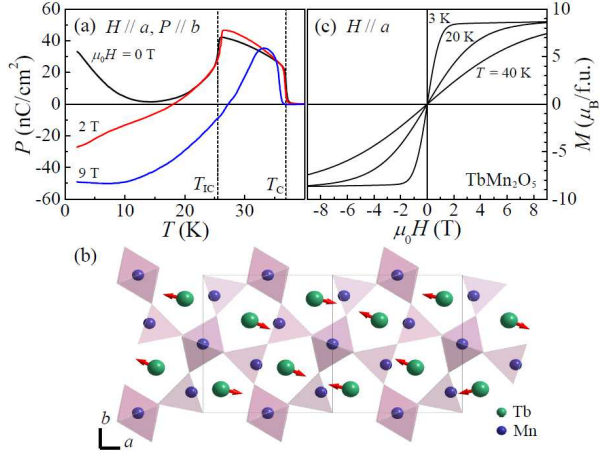}
\end{center}
\caption{(color online) (a) Temperature dependence of electric polarization ($P$) at selected magnetic fields. $T_{\rm C}$ and $T_{\rm IC}$ refer to the commensurate and incommensurate antiferromagnetic transition temperatures. (b) Tb spin configuration of TbMn$_{2}$O$_{5}$ at 20 K reproduced from Ref. 4. (c) Magnetization ($M$) along the $a$-axis at 3, 20, and 40 K.}
\label{fig1}
\end{figure}

Upon cooling, TbMn$_{2}$O$_{5}$ passes through three main magnetic and electric transitions: an incommensurate (ICM) magnetic ordering at $T_{\rm N}\approx43$ K, and a nearly concomitant ferroelectric and commensurate (CM) magnetic ordering at $T_{\rm C}\approx37$ K, and a reentrant low temperature incommensurate (LT-ICM) magnetic ordering at $T_{\rm IC}\approx25$ K with a sharp decrease in $P_{b}$ (See, Fig. 1(a)).\cite{chapon1, nhur} Those transitions are also accompanied by structural anomalies.\cite{cruz} The ferroelectricity and the structural instability is postulated to stem from atomic displacements of Mn$^{3+}$ ions located at the centers of bipyramids.\cite{chapon1,chapon2} Although the antiparallel alignment of Tb spin moments, as shown in Fig. 1(b), has been extracted from the neutron scattering refinement below $\sim$20 K,\cite{chapon1,blake} there is no clear evidence for a thermal transition of Tb spin ordering below $T_{\rm C}$, in contrast to the case of Dy spins in DyMn$_{2}$O$_{5}$.\cite{blake} Moreover, the three thermal transition of TbMn$_{2}$O$_{5}$ are quite similar to those of an isostructural YMn$_{2}$O$_{5}$ without any rare earth ion.\cite{kagomiya} Owing to these facts, the effect of Tb$^{3+}$ ions on the physical properties of TbMn$_{2}$O$_{5}$ appears small.

However, there exist a couple of experimental features that warrant explanations based on the Tb spin effects on TbMn$_{2}$O$_{5}$. First, in contrast to YMn$_{2}$O$_{5}$, in which the negative $P_{b}$ hardly changes up to 9 T in the LT-ICM phase,\cite{haam1} $P_{b}$ of TbMn$_{2}$O$_{5}$ increases with decreasing temperature below $\sim$15 K at $\mu_{0}H=0$ T, and this low temperature positive $P_{b}$ is drastically suppressed to become negative at $\mu_{0}H=2$ T and even more at 9 T.\cite{nhur} Second, the temperature range for the $P_{b}$ increase is consistent with that of the Tb moment increase observed by neutron scattering, thereby indicating a nontrivial coupling between $P_{b}$ and Tb spins.\cite{chapon1} Third, as shown in the isothermal $M$ vs. $H$ curves in Fig. 1(c), the spins of Tb$^{3+}$ ions align within $\mu_{0}H\sim2.5$ T at 3 K and $\sim8$ T even at 20 K to nearly reach a predicted saturated moment ($M_{\rm s}$) of $9\mu_{\rm B}/$f.u. ($4f^{8}$, $^{7}F_{6}$) and thus, the Tb spin alignment is a dominant source of $M$.\cite{haam2}

\begin{figure}
\begin{center}
\includegraphics[width=0.48\textwidth]{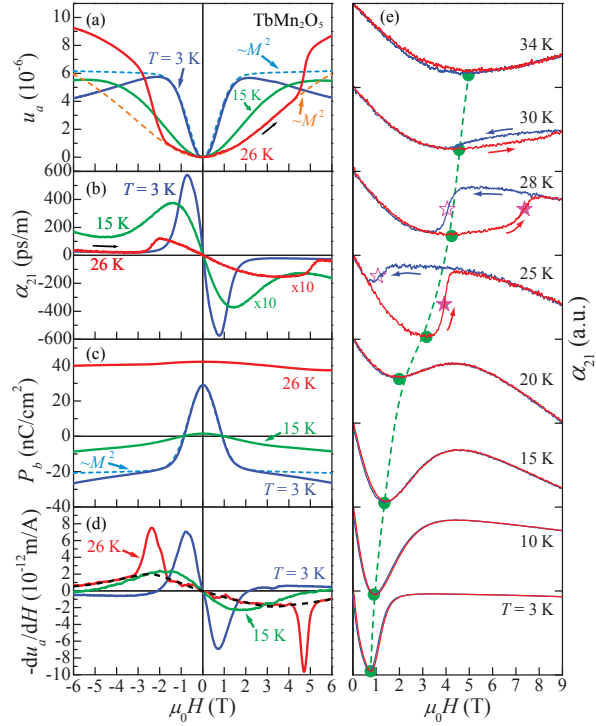}
\end{center}
\caption{(color online) $H$-dependence of (a) longitudinal magnetostriction $u_{a}$, (b) magnetoelectric susceptibility $\alpha_{21}=\delta P_{b}/\delta H_{a}$, (c) $P_{b}$ determined from the integration of $\alpha_{21}$ with $H$, and (d) -d$u_{a}$/d$H$ at 3, 15, and 26 K. The dashed lines in (a) and (c) represent scaled $M^{2}$ curves at 3 and 26 K to fit into the low field. A dashed line in (d) represents a guide to eye to illustrate the asymmetric line shape of -d$u_{a}$/d$H$ at 26 K. (e) Enlarged $\alpha_{21}(H)$ at positive $H$ region at various temperatures. Dashed line is a guide to eye.}
\label{fig2}
\end{figure}

The large $M$ due to the Tb spin alignment results in a significant change in length, under $H$, i.e., magnetostriction in TbMn$_{2}$O$_{5}$. Figure 2(a) shows that the $u_{a}$ is positive and increases in proportion to $M^{2}$. The $u_{a}$ value of $+6\times10^{-6}$ at 2 T is indeed similar to the longitudinal magnetostriction observed in compounds with the Tb$^{3+}$ ions; for example, longitudinal striction is $+2\times10^{-5}$ in TbAlO$_{3}$ at 4 T and $+5\times10^{-5}$ in Tb$_{3}$Ga$_{5}$O$_{12}$ at 2.2 T.\cite{kadomtseva,belov} According to these two features in $u_{a}$, it is most likely that the magnetostriction of TbMn$_{2}$O$_{5}$ is mainly attributed to Tb$^{3+}$ ions involving both single as well as two ion effects as in the case of TbAlO$_{3}$.\cite{kadomtseva}

$M$, $u_{a}$, and $P$ variation under $H$ is closely linked to the Tb spin moment. The $\alpha_{21}(H)$ curves in Fig. 2(b) directly show an evidence for such nontrivial effects of Tb spin moment on ME phenomena. $\alpha_{21}(H)$ at 3 K displays a sharp dip and peak structure around $\pm$0.6 T. Upon being integrated with $H$ as $P_{b}(H)=P_{b}(0)+\int^{H}_{0} \alpha_{21}dH$, $P_{b}(H)$ at 3 K steeply decreases within $|H| <2$ T (Fig. 2(c)), which is consistent with the reported data from pyroelectric current measurements.\cite{nhur,nakamura} The decreasing $P_{b}(H)$ turns out to be proportional to $M^{2}$ in a low $H$ region as is the increase in $u_{a}$, thereby establishing an unambiguous and close correlation between the decrease in $P_{b}$ and increase in $u_{a}$ at 3 K. This correlation is further corroborated by the close similarity in the characteristic asymmetric line shape observed in both -d$u_{a}$/d$H$ and $\alpha_{21}(H)$ curves. Although the absolute value of $u_{a}$ is too small to directly account for the absolute change of $P_{b}$, this correlation reflects that Tb-O distribution can be changed by a local strain of Tb$^{3+}$ ions,\cite{tyson} or exchange interaction between Mn and Tb ions further modulate spin ordering patterns of Mn$^{3+}$ ions\cite{jkoo,wratcliff,lottermoser} to amplify the concomitant $P_{b}$ decrease under $H$. All these observations consistently support that the ME phenomena of TbMn$_{2}$O$_{5}$ at 3 K are coupled with magnetostriction mainly due to the Tb$^{3+}$ ion in a nontrivial way.

It is further noteworthy in Fig. 2 that the magnetostriction effects of Tb$^{3+}$ ion seen in -d$u_{a}$/d$H$ and $\alpha_{21}(H)$ curves are well maintained up to high temperatures. Except the large peaks in the -d$u_{a}$/d$H$ due to the Mn spin transition from the LT-ICM to CM states at 26 K, the asymmetric line shape of the -d$u_{a}$/d$H$ curves is clearly observable at 15 and 26 K (Fig. 2(d)), signaling a significant magnetostriction effect in the entire ferroelectric phases. The $\alpha_{21}(H)$ curves at 15 and 26 K (Fig. 2(b)) also show the characteristic asymmetric line shape, except jumps at 26 K that are coming from the same Mn spin transition. Similar to the relationship between $P_{b}$ and $M^{2}$, the asymmetric line shape of -d$u_{a}$/d$H$ results in the characteristic increase in $u_{a}$ proportional to $M^{2}$, thereby demonstrating the nontrivial coupling between $P_{b}$ and $u_{a}$ at temperatures below $T_{\rm C}$.

\begin{figure}
\begin{center}
\includegraphics[width=0.48\textwidth]{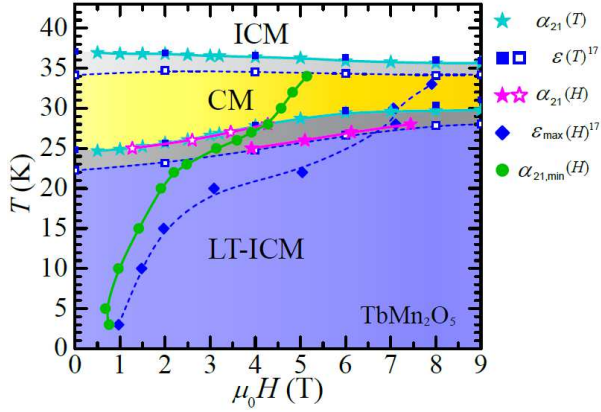}
\end{center}
\caption{(color online) Temperature ($T$) vs. magnetic field ($H$) phase diagram of TbMn$_{2}$O$_{5}$. Asterisks and squares indicate the phase boundaries determined by the $\alpha$ and $\epsilon$ measurements, respectively. Solid and open symbols represent the data measured during the $H$- or $T$-increasing and decreasing runs, respectively. Solid circles and diamonds refer to the points of $\alpha_{21}(H)$ minima and $\epsilon(H)$\cite{haam2} maxima, respectively.} \label{fig3}
\end{figure}

To estimate the phase region affected by Tb$^{3+}$ magnetostriction, we trace the characteristic minimum positions, $\alpha_{\rm 21,min}(H)$, seen in the asymmetric line shape of $\alpha_{21}(H)$ at $H>0$ (solid circles in Fig. 2(e)) and plotted in the phase diagram of Fig. 3. The $\alpha_{\rm 21,min}(H)$ exist at all temperature regions below $T_{\rm C}$. The phase boundaries for the LT-ICM to CM transitions of Mn spins are also determined from the hysteretic jumps in the $\alpha_{21}(H)$ (asterisks in Fig. 2(e)), $\alpha_{21}(T)$ curves (not shown here), and in our previously published $\epsilon(T)$ data\cite{haam2}. In the CM phase region, the trace of the $\alpha_{\rm 21,min}(H)$ (solid circles) is significantly shifted to higher fields, thereby indicating that the complete alignment of Tb spins becomes easier at low temperatures due to the increment of thermal entropy in the LT-ICM phase. $\epsilon(H)$ showed a maximum, $\epsilon_{\rm max}(H)$, of which trace was determined from the results in Ref. 17 (diamonds in Fig. 3). Although $\epsilon_{\rm max}(H)$ is shown at somewhat larger $H$, it shows a similar curvature change as the trace of $\alpha_{\rm 21,min}(H)$, thereby indicating that Tb magnetostriction also affects the magnetodielectric effect.

\begin{figure}
\begin{center}
\includegraphics[width=0.48\textwidth]{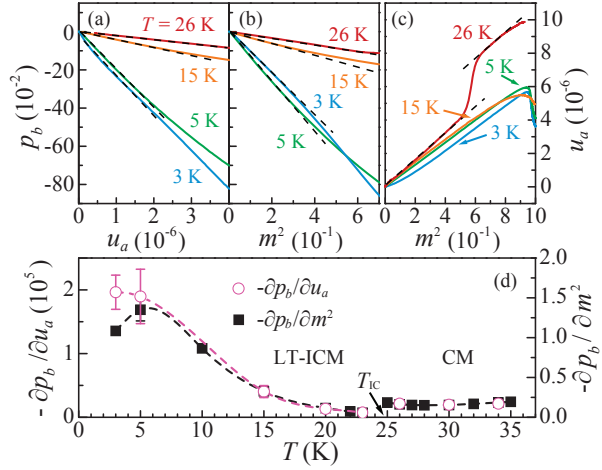}
\end{center}
\caption{(color online) Plots of (a) $p_{b}$ vs. $u_{a}$, (b) $p_{b}$ vs. $m^{2}$, and (c) $u_{a}$ vs. $m^{2}$ at 3, 5, 15, and 26 K. Dashed lines are linear fit lines for low field region. (d) Comparison of the temperature-dependence of $-\partial p_{b}/\partial u_{a}$ (left, circles) and $-\partial p_{b}/\partial m^{2}$ (right, solid squares).} \label{fig3}
\end{figure}

We further uncover that the isothermal variation of $P_{b}$, $u_{a}$, and $M^{2}$ in a low field region roughly follows a simple relationship, i.e., $P_{b}\propto u_{a}$ and $M^{2}$. Figure 4(a)-(c) shows a comparison of three unitless quantities $p_{b}$, $u_{a}$, and $m^{2}$. Here, for the convenience of description, we define $p_{b} \equiv P_{b}/P_{\rm max}$ and $m \equiv M/M_{\rm s}$ with $P_{\rm max} =$ 42.3 nC/cm$^{2}$ ($P$ at 26 K and 0 T) and $M_{\rm s} = 9 \mu_{\rm B}/$f.u. As shown in the figures, the variation of $p_{b}$, $u_{a}$, and $m^{2}$ is roughly linear to each other, except a jump due to the Mn spin transition. We further note in Fig. 4 that there exists characteristic temperature-dependence in their linear relationship; the linear slopes of $p_{b}$ vs. $u_{a}$ and $p_{b}$ vs. $m^{2}$ curves show strong temperature dependence while those of $u_{a}$ vs. $m^{2}$ curve is nearly temperature-independent. The overlapping of $-\partial p_{b}/\partial u_{a}$ and $-\partial p_{b}/\partial m^{2}$ curves at all temperatures below $T_{\rm C}$ with a single constant multiplication (Fig. 4(d)) confirms that the slope changes in both $p_{b}$ vs. $u_{a}$ and $p_{b}$ vs. $m^{2}$ curves follow almost the same temperature-dependence.

To understand the intriguing coupling among $P(=P_{b})$, $M$, and $u(=u_{a})$, we consider a free energy that effectively considers the magnetoelastic effect of Tb$^{3+}$ ions.
\begin{equation}\label{1}
    F_{H}(P,M,u)=\frac{{(P-P_{H=0})}^{2}}{2\chi_{e,T}}+\frac{M^{2}}{2\chi_{m,T}}+\frac{1}{2}C_{T}u^{2}-\frac{\lambda}{2}P^{2}M^{2}-\frac{f}{2}P^{2}u-gM^{2}u-MH
\end{equation}

The first three terms describe temperature-dependence of the order parameters $P$, $M$, and $u$. To describe the variation of quantities under low $H$-regime, we assume that $P_{H=0}$, $\chi_{e,T}$, $\chi_{m,T}$, and $C_{T}$ have predetermined temperature-dependence, consistent with the experimental data. The temperature-dependent evolution of $P_{H=0}$ and $\chi_{e,T}$ has been well studied and appears to be mainly determined by the Mn spin interactions,\cite{chapon1,chapon2,jokamoto,jkoo} while $\chi_{m}$ can be determined by Mn-Mn, Tb-Tb, and Tb-Mn interactions.\cite{chapon1} However, to our knowledge, elastic stiffness constant ($C_{T}$) has not been known yet. The next three terms describe couplings among the order parameters. These are invariant with both inversion and time reversal symmetry operations. Here, $\lambda$, $f$, and $g$ correspond to the temperature-independent coupling constants that are specific to the material.

By minimizing Eq. (1) with $M$, the usual form of $M=\chi_{m,T}H$ is obtained under a condition of $1/\chi_{m,T} \gg{(2gu+\lambda P^{2})}$. With the replacement of $M$ with $\chi_{m,T}H$, $F$ becomes a function of $P$ and $u$. The simultaneous minimization of $F$ with respect to $P$ and $u$ further provides two linear equations with variables $P$ and $u$. By obtaining the functional form of $\partial P/\partial H^{2}$ and $\partial u/\partial H^{2}$ from the two linear equations, one can represent $\partial u/\partial M^{2}$ and $\partial P/\partial M^{2}$, as shown below.
\begin{equation}\label{2}
\frac{\partial u}{\chi^{2}_{m,T}\partial H^{2}}=\frac{(\frac{1}{\chi_{e,T}}-fu-\lambda\chi^{2}_{m,T}H^{2})g+f\lambda P^{2}}{(\frac{1}{\chi_{e,T}}-fu-\lambda\chi^{2}_{m,T}H^{2})C_{T}-f^{2}P^{2}}=\frac{\partial u}{\partial M^{2}}
\end{equation}
\begin{equation}\label{3}
\frac{\partial P}{\chi^{2}_{m,T}\partial H^{2}}=\frac{fgP+\lambda C_{T}P}{(\frac{1}{\chi_{e,T}}-fu-\lambda\chi^{2}_{m,T}H^{2})C_{T}-f^{2}P^{2}}=\frac{\partial P}{\partial M^{2}}
\end{equation}
According to the results shown in Figs. 4 (b) and (c), $\partial u/\partial M^{2}$ is temperature-independent, while $\partial P/\partial M^{2}$ is not. To satisfy both of these constraints, the second term in the denominator might be dominant over the first, which then results in $\partial u/\partial M^{2}=-\lambda/f$, $\partial P/\partial M^{2}=-(fg + \lambda C_{T})/f^{2}P$, and finally, $\partial P/\partial u=(fg + \lambda C_{T})/f\lambda P$. In this approximation, Eq. (1) can successfully explain the temperature-independence of $\partial u/\partial M^{2}$ as well as the same temperature-dependent variation of $\partial P/\partial M^{2}$ and $\partial P/\partial u$ since the $C_{T}$ and $P$ commonly determine the temperature dependence of the last variables. Therefore, our approach of an effective free energy, which is based on the magnetoelastic effect of Tb spins, can provide explanations on the intriguing coupling relationship among $P$, $u$, and $M$.

All above experimental results and considerations based on the free energy constitute compelling evidences for the existence of significant magnetoelastic effect due to Tb$^{3+}$ ion in TbMn$_{2}$O$_{5}$ so as to modulate the macroscopic physical quantities of $M$, $P$, and $u$, which correspond to the spin, charge, and lattice degrees of freedom in TbMn$_{2}$O$_{5}$, respectively. The ME phenomena of TbMn$_{2}$O$_{5}$ such as the actuation of $P$ and the variation of $P$ proportional to $M^{2}$ can only be explained by the magnetoelastic effects of Tb spin moment. Our results provide several implications for the physics of TbMn$_{2}$O$_{5}$ and related multiferroics. First, temperature-dependent elastic constant $C_{T}$ can be directly linked to the temperature-dependent variation of $\partial P/\partial M^{2}$ and $\partial P/\partial u$ via above Eqs. (1)-(3). This scenario can be checked from a direct measurement of $C_{T}$. Second, a microscopic mechanism of how the Tb spin alignment with $H$ can amplify the $P$ variation is a subject worthy of further exploration. The issue can be linked to either a local strain field of Tb magnetostriction or direct exchange coupling between Mn and Tb spins. Third, for a proper description of multiferroic phenomena as well as its application, consideration of the magnetoelastic effects of magnetic ions can be generally important in many other multiferroic materials as well.\cite{tokunaga}

In conclusion, we have provided clear experimental evidences and theoretical indications that magnetostriction due to the Tb spin alignment crucially affects the ME phenomena of TbMn$_{2}$O$_{5}$ in the entire ferroelectric phase. Our results imply that a proper control of the strain or magnetic moment of rare earth ions can be useful in the application of existing multiferroics in a low field phase.

We thank Maxim Mostovoy for fruitful discussions on the free-energy analysis. This study was supported by NRF through Creative Research Initiatives, NRL (M10600000238) , Basic Science Research (2009-0083512) programs and by GPP program (K20702020014-07E0200-01410). GSJ was also supported by Basic Science Research Program (2010-0010937). The work at Rutgers was supported by the DOE grant (DE-FG02-07ER46382). VFC is a member of CONICET.


\begin{thebibliography}{21}
\bibitem{chapon1} L. C. Chapon, {\it et al}., Phys. Rev. Lett. {\bf 93}, 177402 (2004).
\bibitem{chapon2} L. C. Chapon, P. G. Radaelli, G. R. Blake, S. Park, and S.-W. Cheong, Phys. Rev. Lett. {\bf 96}, 097601 (2006).
\bibitem{kagomiya} I. Kagomiya, {\it et al}., Ferroelectrics {\bf 286}, 167 (2003).
\bibitem{blake} G. R. Blake, {\it et al}., Phys. Rev. B {\bf 71}, 214402 (2005).
\bibitem{jwkim} J. W. Kim {\it et al}., Proc. Natl. Acad. Sci. U. S. A. {\bf 106}, 15573 (2009); Gun Sang Jeon, Jin-Hong Park, Jae Wook Kim, Kee Hoon Kim, and Jung Hoon Han, Phys. Rev. B {\bf 79}, 104437 (2009).
\bibitem{hkimura} H. Kimura, {\it et al}., J. Phys. Soc. Jpn. {\bf 76}, 074706 (2007).
\bibitem{mfukunaga} M. Fukunaga {\it et al}., Phys. Rev. Lett. {\bf 103}, 077204 (2009).
\bibitem{jokamoto} J. Okamoto, {\it et al}., Phys. Rev. Lett. {\bf 98}, 157202 (2007).
\bibitem{jkoo} J. Koo {\it et al}., Phys. Rev. Lett. {\bf 99}, 197601 (2007).
\bibitem{wratcliff} W. Ratcliff, {\it et al}., Phys. Rev. B {\bf 72}, 060407(R) (2005).
\bibitem{tyson} T. A. Tyson, M. Deleon, S. Yoong, and S.-W. Cheong, Phys. Rev. B {\bf 75}, 174413 (2007).
\bibitem{lottermoser} T. Lottermoser, D. Meier, R. V. Pisarev, and M. Fiebig, Phys. Rev. B {\bf 80}, 100101(R) (2009).
\bibitem{nhur} N. Hur, {\it et al}., Nature {\bf 429}, 392 (2004).
\bibitem{hryu} H. Ryu {\it et al}., Appl. Phys. Lett. {\bf 89}, 102907 (2006); Y. S. Oh, {\it et al}., Appl. Phys. Lett. {\bf 97}, 052902 (2010).
\bibitem{cruz} C. R. dela Cruz, {\it et al}., Phys. Rev. B {\bf 73}, 100406(R) (2006).
\bibitem{haam1} S. Y. Haam {\it et al}. unpublished.
\bibitem{haam2} S. Y. Haam, {\it et al}., Ferroelectrics {\bf 336}, 153 (2006).
\bibitem{kadomtseva} A. M. Kadomtseva, {\it et al}., J. Magn. Magn. Mater. {\bf 81}, 196 (1989).
\bibitem{belov} K. P. Belov {\it et al}., JETP Letters {\bf 4}, 127 (1966).
\bibitem{nakamura} H. Nakamura, {\it et al}., Ferroelectrics {\bf 204}, 107 (1997).
\bibitem{tokunaga} Y. Tokunaga, {\it et al}., Nature Materials {\bf 8}, 558 (2009).
\end{thebibliography}
\end{document}